\documentclass[journal]{IEEEtran}
\usepackage{mathrsfs}
\usepackage{tabu}
\usepackage{bbm}
\usepackage{color}
\usepackage{amsmath,amscd,amssymb,latexsym}
\usepackage[all]{xy}
\usepackage{pgf,tikz}

\newtheorem{theorem}{Theorem}[section]
\newtheorem{lemma}[theorem]{Lemma}
\newtheorem{corollary}[theorem]{Corollary}
\newtheorem{definition}[theorem]{Definition}
\newtheorem{example}[theorem]{Example}

\newtheorem{remk}[theorem]{Remark}

\ifCLASSINFOpdf
\else
\fi
\begin{document}
%
\title{ Binary LCD Codes and Self-orthogonal Codes via Simplicial Complexes }
%
%
%

\author{Yansheng Wu and
     Yoonjin Lee 
     
     \thanks{Manuscript received December 30, 2019; accepted March 16, 2020. The paper is supported by Basic Science Research Program through the National Research Foundation of Korea (NRF) funded by the Ministry of Education (Grant No. 2019R1A6A1A11051177) and also by the National Research Foundation of Korea (NRF) grant funded by the Korea government (MEST)(NRF-2017R1A2B2004574). The associate editor coordinating the review of this letter and approving it for publication was Marco Baldi. (Corresponding author: Yoonjin Lee.)}
\thanks{Yansheng Wu and Yoonjin Lee are both with the Department of Mathematics, Ewha Womans University, Seoul 03760, South
Korea (email: wysasd@163.com; yoonjinl@ewha.ac.kr)
}

\thanks{Digital Object Identifier ~~~~~~~~~}
}

\maketitle

\begin{abstract} Due to some practical applications, linear complementary
dual (LCD) codes and self-orthogonal codes have attracted wide attention in recent years.
In this paper, we use simplicial complexes for construction of an infinite family  of
binary LCD codes and two infinite families of binary self-orthogonal codes.
Moreover, we explicitly determine the weight distributions of these codes.
We obtain binary LCD codes which have minimum weights two or three,
and we also find some self-orthogonal codes meeting the Griesmer bound.
As examples, we also present some (almost) {\it optimal} binary self-orthogonal codes and
LCD {\it distance optimal} codes.
\end{abstract}


%
\IEEEpeerreviewmaketitle

\section{Introduction}
%
%
%
%

\IEEEPARstart{T}{he} concept of linear complementary dual (LCD) codes was introduced by
Massey \cite{M} in 1992. 
For implementations against side-channel and fault injection attacks, 
a new application of binary LCD codes was found by Carlet and Guilley (\cite{BCCGM, CG}).
Since then, LCD codes have attracted wide attention from the coding research community~(\cite{CMTQ1, CMTQP}, \cite{ KY}-\cite{LDL},  \cite{Y,ZTLD}).
Carlet {\em et al.} \cite{CMTQP} proved that for $q >3$, any $q$-ary linear code  is equivalent to an LCD code over $\Bbb F_q$;
therefore, it is sufficient to investigate binary LCD codes and ternary LCD codes.
Self-orthogonal codes are very important for the study of quantum communications and quantum computations
since they can be applied to the classical construction of quantum error-correcting codes (\cite{CRSS1, CRSS2}).

In this paper, we mainly use simplicial complexes for constructing binary LCD codes and binary self-orthogonal codes.
For the definition of simplicial complexes, we need the following notations. Let $\Bbb F_2$ be the finite field of order $2$ and
$m$ be a positive integer. The support $\mathrm{supp}(v)$ of a vector $v$ in $\Bbb F_2^m$ is defined by the set of nonzero coordinate positions of $v$.
Let $2^{[m]}$ denote the power set of $[m]=\{1, \ldots, m\}$. It is easy to check that there is a bijection between $\mathbb{F}_2^m$ and $2^{[m]}$, defined by $v\mapsto$ supp$(v)$; hence, due to this bijection, a vector $ v$ in $\mathbb{F}_2^m$ is identified with its support supp$(v)$.
For two sets $A$ and $B$,
the set $\{x: x\in A\mbox{ and } x\notin B\}$ is denoted by $A\backslash B$,  and the size of  $A$ is denoted by  $|A|$.

\begin{definition}
A subset $\Delta$ of $\mathbb{F}_2^m $ is called a {\it simplicial complex} if $u\in \Delta$ and $\mathrm{supp}(v)\subseteq \mathrm{supp}(u)$ imply $v\in \Delta$ for any $u,v\in \Bbb F_2^m$.
\end{definition}

An element of a simplicial complex $\Delta$ is called \emph{maximal} if it is not properly contained in the others in $\Delta$.
Let $\mathcal{F}$ be the set of maximal elements of a simplicial complex $\Delta$. Especially,  $\Delta_F$ denotes the simplicial complex generated by a nonzero vector $F$ in $\Bbb F_2^m$.


In this paper, we use a typical construction of a linear code given in \cite{HP}.
Let $D=\{ g_1, g_2,\ldots, g_n \}\subseteq \Bbb F_p^m$. Then a linear code $\mathcal{C}_{D}$ of length $n=|D|$ over $\Bbb F_{p}$ can be defined by
\begin{equation}\mathcal{C}_{D}= \{c_{u}=(u\cdot g_1, u\cdot  g_2, \ldots, u\cdot { g}_n): { u}\in  \Bbb F_{p}^m\}, \end{equation}  where $\cdot $ denotes the Euclidean inner product of two elements in $\Bbb F_p^m.$ The set $D$ is  called the {\it defining set} of  $\mathcal{C}_{D}$. Let $G$ be the $m\times n$ matrix as follows:
\begin{equation}G=[g_1^T \; g_2^T \; \cdots \; g_n^T],\end{equation}
where the column vector $g_i^T$ denotes the transpose of a row vector $g_i$.
Zhou {\em et al.} \cite{ZTLD} obtained some simple conditions under which the linear codes defined in Eq. (1) are LCD or self-orthogonal, and they also presented four infinite families of binary linear codes. For any positive integers $m$ and $t$ with $1\le t\le m-1$, two defining sets are given as follows:
$$D_t=\{g\in \Bbb F_2^m: wt(g)=t\}, $$
$$\mbox{ and } D_{\le t}=\{g\in \Bbb F_2^m: 1\le wt(g)\le t\},$$ where $wt(v)$ denotes the Hamming weight  of $v\in \mathbb{F}^m_2$.
We note that the two sets can also be expressed by using simplicial complexes in the following way:
$$D_t= \Delta_{D_t}  \setminus  \Delta_{D_{t-1}}  \mbox{ and } D_{\le t}= \Delta_{D_t}  \backslash\{0\}.$$
Note that here  ${D_t} $ denotes a set of maximal elements for any $t\ge 1$, and
$\Delta_{D_t}$ and $\Delta_{D_{t-1}}$ are simplicial complexes. For example, if $m=3$, then $D_1=\{(1,0,0), (0,1,0), (0,0,1)\}$ and $D_2=\{(1,1,0), (1,0,1),(0,1,1)\}$; hence, $\Delta_{D_1}=D_1\cup \{0\}$ and $\Delta_{D_2}=  \{0\}\cup D_1\cup D_2$. It is easy to check that $\Delta_{D_1}$ and $\Delta_{D_2}$  are simplicial complexes.

Inspired by \cite{ZTLD}, we employ the difference of two distinct simplicial complexes
for construction of an infinite family  of
binary LCD codes and two infinite families of binary self-orthogonal codes.
This paper is organized as follows.
In Section II we introduce some basic concepts on generating functions, LCD codes, and self-orthogonal codes.
In Section III  we determine the weight distributions of some  binary linear codes and discuss the minimum distances of their dual codes.
Section IV presents a class of binary LCD codes and two classes of binary self-orthogonal codes. Section V concludes this work.

\section{Preliminaries}

\subsection{ Generating functions }


The following $m$-variable generating function associated with a subset $X$ of  $\mathbb{F}_2^m$ was introduced by Chang {\em et al.}  \cite{CH}.
$$\mathcal{H}_{X}(x_1,x_2\ldots, x_m)=\sum_{u\in X}\prod_{i=1}^mx_i^{u_i}\in \mathbb{Z}[x_1,x_2, \ldots, x_m],
$$
where $u=(u_1,u_2,\ldots, u_m)\in \mathbb{F}_2^m$; here, for $u_i$, we use the identification of $0, 1 \in \mathbb{F}_2$ with $0, 1 \in \mathbb{Z}_2$, respectively (Note that this is just a formal definition and there should be no confusion because
we do not make addition operation on the powers of $x_i$).

The following lemma will be used in Section III.

\begin{lemma}\cite[Theorem 1]{CH}  \label{th1}
Suppose that  $\Delta$ is a simplicial complex of $\mathbb{F}_2^m$ and $\mathcal {F}$ is  the set of maximal elements of $\Delta$. Then
 \begin{align*}
 \mathcal{H}_{\Delta}(x_1,x_2\ldots, x_m)=\sum_{\emptyset\neq S\subseteq \mathcal{F}}(-1)^{|S|+1}\prod_{i\in \cap S}(1+x_i).
 \end{align*}
\end{lemma}

\begin{remk} Recall that there is a bijection between $\mathbb{F}_2^m$ and $2^{[m]}$. 
Hence, the set $\cap S$ in Lemma 2.1 can be understood as the intersection of the elements of $S$ in $2^{[m]}$. We have the following example. Let $\Delta=\langle (1,1,0), (0,1,1) \rangle$ be a simplicial complex in $\Bbb F_2^3$. By Lemma 2.1, we have  \begin{eqnarray*}&&\mathcal{H}_{\Delta}(x_1,x_2, x_3)\\
&=&(1+x_1)(1+x_2)+(1+x_2)(1+x_3)-(1+x_2)\\
&=&1+x_1+x_2+x_3+x_1x_2+x_2x_3.\end{eqnarray*}
\end{remk}

\subsection{  LCD codes and self-orthogonal codes }

Let $\Bbb F_q$ be the finite field of order $q$, where $q$ is a power of a prime.  
Let $\mathcal C$ be an $[n,k]$ code over $\Bbb F_q$.
The dual code $\mathcal{C}^{\bot}$ of $\mathcal{C}$ is defined by $\mathcal{C}^{\bot}=\{w\in\Bbb F_q^n: w\cdot c=0 \mbox{ for every } c\in \mathcal{C}\}.$ If $\mathcal{C} \cap \mathcal{C}^{\bot}=\{0\}$, then   $\mathcal{C}$ is called a  {\it linear complementary dual} (LCD) code; if $\mathcal{C} \subseteq\mathcal{C}^{\bot}$, then $\mathcal{C}$ is called  {\it self-orthogonal}.

Regarding the codes defined in Eq. (1), Zhou {\em et al.} \cite{ZTLD} obtained the following lemma.

\begin{lemma} \cite[Corollary 16]{ZTLD}
Let $\mathcal{C}_D$ be the linear code defined in Eq. (1).
Let $\mbox{Rank}(G)$ denote the rank of the matrix $G$ in Eq. (2).
Then $\mathcal{C}_D$ is self-orthogonal (LCD, respectively) if and only 
if $GG^T=0$  ($\mbox{Rank} (GG^T)=\mbox{Rank}(G)$, respectively).
\end{lemma}

Let $\mathcal{C}$ be an $[n,k,d]$ linear code over $\Bbb F_q$.
Assume that there are $A_i$ codewords in $\mathcal C$ with Hamming weight $i$ for $1\le i \le n$.
Then $\mathcal C$ has weight distribution $(1, A_1, \ldots, A_n)$ and  weight enumerator $1+A_1z+\cdots +A_nz^n$.
 Moreover, if the number of nonzero $A_{i}$'s in the sequence $(A_1, \ldots, A_n)$ is exactly equal to $t$,
 then the code is called {\it $t$-weight}. An $[n,k,d]$ code  $\mathcal{C}$ is called {\it distance optimal} if there is no $[n,k,d+1]$ code (that is, this code has the largest minimum distance for given length $n$ and dimension $k$),
 and it is called {\it almost optimal} if an $[n, k, d + 1]$ code is  distance optimal (refer to~\cite[Chapter 2]{HP}). On the other hand, the {\it Griesmer bound} \cite{G} on an $[n, k, d]$  linear code over $\Bbb F_q$ is given by
 $ \sum_{i=0}^{k-1}\bigg\lceil {\frac{d}{q^i}} \bigg\rceil \le n,  $ where $\lceil {\cdot} \rceil$ is the ceiling function.

Furthermore, a binary $[n,k,d]$ LCD code $\mathcal C$ is called {\it LCD distance optimal} if there is no $[n,k,d+1]$ LCD code (that is, this LCD code has the largest minimum distance among $[n, k]$ LCD codes for given length $n$ and dimension $k$),
 and it is called {\it LCD almost optimal} if an $[n, k, d + 1]$ code is LCD distance optimal.

\section{ Weight distributions of binary linear codes arising from simplicial complexes}

We will determine the weight distributions of the codes defined in Eq. (1), noting that their defining sets are
expressed as the differences of two simplicial complexes.

Let $\Delta_1$ and $\Delta_2$ with $\Delta_2 \subset \Delta_1$ be two distinct simplicial complexes of $\Bbb F_2^m$.
Let $p=2$ and $D=\Delta_1\backslash \Delta_2$ in Eq. (1). Note that if $u=0$ in Eq. (1), then $wt(c_{u})=0$.
From now on, we assume that $u\neq {0}$. Then
\begin{eqnarray}wt(c_{u})
&=&|D|-\frac 12 \sum_{y\in \Bbb F_2}\sum_{d\in D}(-1)^{y(u\cdot d)}\nonumber\\
&=& \frac{|D|}2-\frac 12 \sum_{d\in \Delta_1\backslash \Delta_2}(-1)^{u\cdot d}\nonumber\\
&=& \frac{|D|}2-\frac 12(\sum_{d\in \Delta_1}(-1)^{u\cdot d}- \sum_{d\in \Delta_2}(-1)^{u\cdot d})\nonumber\\
&=& \frac{|D|}2-\frac 12\mathcal{H}_{\Delta_1}((-1)^{u_1},\ldots, (-1)^{u_m})\nonumber\\
&+&\frac 12\mathcal{H}_{\Delta_2}((-1)^{u_1},\ldots, (-1)^{u_m}),
\end{eqnarray}
where $u=(u_1, u_2, \ldots, u_m)^{}\in \Bbb F_2^m$.

For $u\in \mathbb{F}_2^m$ and  $X\subseteq\mathbb{F}_2^m$,
a Boolean function $\chi(u|X)$ in $m$-variable is defined by
  $\chi(u|X)=1$ if and only if $u\bigcap X=\emptyset$.
If a simplicial complex is generated by a maximal element $A$ (denoted by $\Delta_A$), then by Lemma 2.1 we have
\begin{eqnarray}&&\mathcal{H}_{\Delta_A}((-1)^{u_1}, \ldots, (-1)^{u_m})= \prod _{i\in A}(1+(-1)^{u_i})\nonumber\\
&=&\prod _{i\in A}2(1-u_i)=2^{|A|}\chi(u | A).\end{eqnarray}
\begin{theorem} {\rm Let $m\ge 3$ be a positive integer.
 Suppose that $A$ and $B$ are two  elements of $\Bbb F_2^m$ with $B\subset A$.
 Let $D=\Delta_A \backslash \Delta_B$. Then the code $\mathcal{C}_D$ defined in Eq. (1) meets the Griesmer bound.

 $(1)$ If $|B|=0$, then  $\mathcal{C}_D$ is a $[2^{|A|}-1, |A|, 2^{|A|-1}]$ one-weight code with  weight enumerator $1+(2^{|A|}-1)z^{2^{|A|-1}}.$

$(2)$ If $|B|\ge 1$, then   $\mathcal{C}_D$ is a $[2^{|A|}-2^{|B|}, |A|, 2^{|A|-1}-2^{|B|-1}]$ two-weight code with weight enumerator  $$1+(2^{|A|}-2^{|A|-|B|})z^{2^{|A|-1}-2^{|B|-1}}+(2^{|A|-|B|}-1)z^{2^{|A|-1}}.$$

 }

\end{theorem}

{\bf Proof}  The length of  $\mathcal{C}_D$ is $2^{|A|}-2^{|B|}$.  By  Eqs. (3) and (4), \begin{equation*}wt(c_{{u}})=2^{|A|-1}(1-\chi(u|A))- 2^{|B|-1}(1-\chi(u|B)).
\end{equation*}
The frequency of each codeword $c_{{u}}$ can be determined by the vector $u$. By the definition of $\chi(u|A)$, we note that $wt(c_u)=0$ if and only if $\chi(u|A)=\chi(u|B)=1$: that is, $u\cap A=\emptyset$. Since $u\in \Bbb F_2^m$, every codeword is repeated $2^{m-|A|}$ times. Hence, we see that
the code $\mathcal{C}_D$ has dimension $|A|$.

Furthermore, if $|B|\ge 1$, then we have \begin{eqnarray*} &&\sum_{i=0}^{|A|-1}\bigg\lceil {\frac{2^{|A|-1}-2^{|B|-1}}{2^i}} \bigg\rceil \\
&=& (2^{|A|}-1)-(2^{|B|}-1) =2^{|A|}-2^{|B|}. \end{eqnarray*} Hence, we conclude that  $\mathcal{C}_D$ meets the Griesmer bound.
Similarly, the result holds for the case where $|B|=0$.
$\blacksquare$

\begin{theorem} Let  $D$ be defined as in Theorem 3.1.  Then $\mathcal{C}_D^{\bot}$ is a
 $[2^{|A|}-2^{|B|}, 2^{|A|}-2^{|B|}-|A|, \delta]$ linear code, where
$$ \delta=\left\{
\begin{array}{ll}
  3   &      \mbox{if}\ |A|>|B|+1,\\
4 & \mbox{if}\ |A|=|B|+1\ge 3.
\end{array} \right. $$
\end{theorem}

{\bf Proof} Assume that $D=\{g_1, \ldots, g_n\}\subseteq \Bbb F_2^m$ with $n=|D|$.
The generator matrix  $G'$ of  $\mathcal{C}_D$ can be induced by the matrix $G$ in Eq. (2) by deleting all the zero row vectors of $G$.
Clearly, $G'$ is the parity-check matrix of  $\mathcal{C}_D^{\bot}$. The minimum distance of $\mathcal{C}_D^{\bot}$ is greater than 2.
We divide the proof into two parts.

(1) If $|A|>|B|+1$, then there are two distinct positive integers $i$ and $j$ in $A\backslash B$.  Let ${\bf e}_k=(e_1, e_2, \ldots, e_m)\in \Bbb F_2^m$, where
$e_k=1$ and $e_l=0 $ if $l\neq k$. Then it is easy to check that ${\bf e}_i^T, {\bf e}_j^T$, and ${\bf e}_i^T+{\bf e}_j^T$ are three different columns of $G'$; therefore, the minimum distance of $\mathcal{C}_D^{\bot}$ is 3.

(2) If $|A|=|B|+1$, then we assume that $A\backslash B=\{i\}$ without loss of generality. We note that any three columns of $G'$ are linearly independent. Since $|B|\ge 2$, there are two integers $j$ and $k$ in $B$. Then ${\bf e}_i^T, {\bf e}_i^T+{\bf e}_j^T$,  ${\bf e}_i^T+{\bf e}_k^T$, and ${\bf e}_i^T+{\bf e}_j^T+{\bf e}_k^T$  are four linearly dependent  columns of $G'$. Therefore, the minimum distance of $\mathcal{C}_D^{\bot}$ is 4.
$\blacksquare$

\begin{corollary} {\rm Let $|B|=0$ and $|A|>1$ in Theorem 3.2. Then  $\mathcal{C}_D^{\bot}$ is a $[2^{|A|}-1, 2^{|A|}-1-|A|, 3]$ Hamming code. }
\end{corollary}

\begin{theorem} {\rm Let $m\ge 3$ be a positive integer.
 Suppose that $A$ and $B$ are two distinct elements of $\Bbb F_2^m$ such that $0< |B|< |A|$ and $A\cap B=\emptyset$.
 Let $D=(\Delta_A\cup \Delta_B) \backslash \{0\}$.
 Then  $\mathcal{C}_D$  in Eq. (1) is a $[2^{|A|}+2^{|B|}-2, |A|+|B|, 2^{|B|-1}]$ three-weight code  with  weight enumerator \begin{eqnarray*}&&1+(2^{|B|}-1)z^{2^{|B|-1}}+(2^{|A|}-1)z^{2^{|A|-1}}\\
 &+&(2^{|B|}-1)(2^{|A|}-1)z^{2^{|B|-1}+2^{|A|-1}}.\end{eqnarray*}

 }

\end{theorem}

{\bf Proof}  The length of  $\mathcal{C}_D$ is $2^{|A|}+2^{|B|}-2$.  By  Eqs. (3) and (4), we have \begin{equation*}wt(c_{{u}})=2^{|A|-1}(1-\chi(u|A))+2^{|B|-1}(1-\chi(u|B)).
\end{equation*}
The frequency of each codeword in $\mathcal{C}_D$ can be determined by the vector $u$, and so the result follows immediately.
$\blacksquare$

In a similar way to Theorem 3.2, we have:

\begin{theorem}
Let  $D$ be defined as in Theorem 3.4.
Then $\mathcal{C}_D^{\bot}$ is a  $[2^{|A|}+2^{|B|}-2, 2^{|A|}+2^{|B|}-2-|A|-|B|, 3]$  code. \end{theorem}

\begin{theorem} {\rm Let $m$ be a positive even integer and $k=\frac m2$.
Let $\{A_1,\ldots, A_k\}$ be a partition of $\{1,2,\ldots,  m\}$, where $|A_i|=2$ for $1\le i\le k$.
Let $D=(\Delta_{A_1}\cup\cdots\cup \Delta_{A_k}) \backslash \{0\}$ in Eq. (1).
Then  $\mathcal{C}_D$ is a $[{3m}/2, m, 2]$ code and its  weight
enumerator is given by $$1+\prod_{l=0}^{k-1}3^l{k\choose l}z^{m-2l}.$$


 }

\end{theorem}

{\bf Proof}  The length of  $\mathcal{C}_D$ is $\frac32m$.  By  Eqs. (3) and (4), \begin{equation}wt(c_{{u}})=m-2(\chi(u|A_1)+\cdots+\chi(u|A_k)).
\end{equation}
The frequency of each codeword in $\mathcal{C}_D$ can be determined by the vector $u$, and hence the result follows right away.
$\blacksquare$

We obtain the following theorem in a similar way to Theorem 3.2.

\begin{theorem} Let $D$ be defined as in Theorem 3.6.
Then $\mathcal{C}_D^{\bot}$ is a  $[{3m}/2, m/2, 3]$  code.
\end{theorem}

\section{ Binary LCD codes and self-orthogonal codes}

We present some binary LCD codes and binary self-orthogonal codes in this section.

\begin{lemma} Let $\Delta_A$ be a simplicial complex generated by a nonzero element $A$ in $\Bbb F_2^m$
and $\Delta_A\backslash\{0\}=\{g_1, g_2,\ldots, g_n\}\subseteq\Bbb F_2^m$, where $n=2^{|A|}-1$.
Let $G=[g_1^Tg_2^T\cdots g_n^T]$ be the $m\times n$ matrix in Eq. (2).

Then $\mbox{Rank}(G)=|A|$ and $$ \mbox{Rank}(GG^T)=\left\{
\begin{array}{ll}
  0   &      \mbox{if}\ |A|\ge 3,\\
|A| & \mbox{if}\ |A|<3.
\end{array} \right. $$

\end{lemma}

{\bf Proof}  Note that $\mbox{Rank}(G)=|A|$.
Let $M=(m_{ij})_{m\times m}=GG^T$. By \cite[Lemma 18]{ZTLD},
assume that $c_i$ is the $i$-th row vector of $G$.
Then $m_{i,j}=c_ic_j^T$.
Let $U_{i,j}=\{g=(g_1,g_2, \ldots, g_m)\in D: g_i=g_j=1\}$.
Then $m_{i,j}=|U_{i,j}| \pmod 2$. Then the result follows from Lemma 2.2 and
$$ U_{i,j}=\left\{
\begin{array}{ll}
  2^{|A|-1}   &   \mbox{if}\  i=j\in A,\\
2^{|A|-2} & \mbox{if}\ i\neq j, \; i,j\in A, \\
0 &\mbox{otherwise}.
\end{array} \right. $$

\begin{theorem} Let $D$ be defined as in Theorem 3.1.
Then the code $\mathcal{C}_D$ defined in Eq. (1) is self-orthogonal if and only if one of the followings holds:

(1) $|B|=0$ and $|A|\ge 3$.

(2) $|A|>|B|\ge 3$.
\end{theorem}

{\bf Proof} Let $\Delta_B\backslash\{0\}=\{g_1, g_2,\ldots, g_l\}\subseteq\Bbb F_2^m$ and $\Delta_A\backslash\Delta_B=\{g_{l+1}, g_{l+2},\ldots, g_n\}\subseteq\Bbb F_2^m$, and
$\Delta_A\backslash\{0\}=\{g_1, g_2,\ldots, g_n\}\subseteq\Bbb F_2^m$.
Let $G_1=[g_1^T g_2^T\cdots g_l^T]$, $G_2=[g_{l+1}^Tg_{l+2}^T\cdots g_n^T]$ and $G=[G_1G_2].$ By Lemma 2.2, the code $\mathcal{C}_D$ is self-orthogonal if and only if $G_2G_2^T=0$. Note that $GG^T=G_1G_1^T+G_2G_2^T$. Now, we consider the following four cases depending on the value of $|B|$.

(1) If $|B|=0$, then $\mathcal{C}_D$ is self-orthogonal if and only if $|A|\ge 3$ from Lemma 4.1.

(2) If $|B|=1$, then $m_{ii}=2^{|A|-1}-1 \equiv1\pmod 2$ for $i\in B$. Hence,  $\mathcal{C}_D$ cannot be self-orthogonal in this case.

(3) If $|B|=2$, then $m_{ij}=2^{|A|-2}-1 \equiv1\pmod 2$ for $i,j\in B$. Thus,  $\mathcal{C}_D$ cannot be self-orthogonal in this case.

(4)  If $|B|\ge3$, then we have that $G_1G_1^T=0$ and $GG^T=0$ by Lemma 4.1. Therefore, $\mathcal{C}_D$ is self-orthogonal.
$\blacksquare$


 \begin{example}{\rm Let $|B|=0$ and $|A|=3\le m$. Then $\mathcal{C}_D$ in Theorem 3.1 is a  $[7,3,4]$ self-orthogonal code, and $\mathcal{C}_D^{\bot}$ is a $[7,4,3]$ code.  According to \cite{G2}, we find that both $\mathcal{C}_D$ and $\mathcal{C}_D^{\bot}$ are distance optimal.}
\end{example}

 \begin{example}{\rm Let $|B|=3$ and $|A|=5\le m$. Then  $\mathcal{C}_D$ in Theorem 3.1 is a $[24,5,12]$ self-orthogonal code, and $\mathcal{C}_D^{\bot}$ is a $[24,19,3]$  code.  We confirm that both $\mathcal{C}_D$ and $\mathcal{C}_D^{\bot}$ are distance optimal according to \cite{G2}.}
\end{example}

 \begin{example}{\rm Let $|B|=4$ and $|A|=5\le m$. Then  $\mathcal{C}_D$ in Theorem 3.1 is a $[16,5,8]$ self-orthogonal code and $\mathcal{C}_D^{\bot}$ is a $[16,11,4]$ code.  According to \cite{G2}, we conclude that both $\mathcal{C}_D$ and $\mathcal{C}_D^{\bot}$ are distance optimal.}
\end{example}

\begin{theorem} Let $D$ be defined as in Theorem  3.4.
Then  $\mathcal{C}_D$ is self-orthogonal if and only if  $|A|>|B|\ge 3$.
\end{theorem}

{\bf Proof} Let $\Delta_B\backslash\{0\}=\{g_1, \ldots, g_l\}\subseteq\Bbb F_2^m$ and
$\Delta_A\backslash\{0\}=\{h_1,\ldots, h_n\}\subseteq\Bbb F_2^m$.
Let $G_1=[g_1^T \cdots g_l^T]$, $G_2=[h_1^T\cdots h_n^T]$, and $G=[G_1G_2].$
From the assumption that $A\cap B=\emptyset$ and Lemma 2.2, it follows that  $\mathcal{C}_D$
is self-orthogonal if and only if $GG^T=0$. The result thus follows from the fact that $GG^T=G_1G_1^T+G_2G_2^T$ and Lemma 4.1.
$\blacksquare$

 \begin{example}{\rm Let $|B|=2$, $|A|=3$, and $5\le m$. Then  $\mathcal{C}_D$ in Theorem 3.4 is  a $[10,5,3]$  self-orthogonal code and $\mathcal{C}_D^{\bot}$ is a $[10,5,3]$ code.  According to \cite{G2},  we find that $\mathcal{C}_D$ and $\mathcal{C}_D^{\bot}$ are both almost optimal.}
\end{example}

\begin{theorem} Let $D$ be defined as in Theorem  3.6. Then  $\mathcal{C}_D$ is an LCD code.
\end{theorem}

{\bf Proof} By Lemma 4.1, for any $1\le i\le k$ we have $m_{i_1,i_2}=m_{i_2,i_1}=1,$ where $\{i_1, i_2\}= A_i$.  Note that $\{A_1,\ldots, A_k\}$ is a partition of $\{1,2,\ldots, m\}$ and $\mbox{Rank}(G)=m$. Equivalently, we can write $GG^T=\mbox{diag}\{I_2, I_2, \ldots, I_2\}$, where $I_2$ is the identity matrix of order 2. Then $\mbox{Rank}(G)=\mbox{Rank}(GG^T)=m$. Then the result follows from  Lemma 2.2.
$\blacksquare$

In \cite{GK, HS}, the authors obtained some bounds on  LCD codes, and they also gave a complete classification of binary LCD codes with small lengths.


 \begin{example}{\rm Let $m=4$. Then $\mathcal{C}_D$ in Theorem 3.6 is a  $[6,4,2]$ binary LCD code and $\mathcal{C}_D^{\bot}$ is a $[6,2,3]$ binary  LCD  code.  According to \cite{G2}, $\mathcal{C}_D$ is distance optimal  and $\mathcal{C}_D^{\bot}$ is almost optimal.  According to the tables in (\cite{GK, HS}), we see that $\mathcal{C}_D$ and $\mathcal{C}_D^{\bot}$ are both LCD distance optimal codes as well.

 }
\end{example}

 \begin{example}{\rm Let $m=6$. Then   $\mathcal{C}_D$ in Theorem 3.6 is a $[9,6,2]$ binary LCD code and $\mathcal{C}_D^{\bot}$ is a $[9,3,3]$ binary LCD code. According to \cite{G2},  $\mathcal{C}_D$ is distance optimal and $\mathcal{C}_D^{\bot}$ is almost optimal.
Moreover, we conclude that the code $\mathcal{C}_D$ is LCD distance optimal based on the tables in (\cite{GK, HS}).

}
\end{example}

 \begin{example}{\rm Let $m=8$. Then  $\mathcal{C}_D$ in Theorem 3.6 is a $[12,8,2]$ binary  LCD code and $\mathcal{C}_D^{\bot}$ is a $[12,4,3]$ binary  LCD code.
 We find that the code $\mathcal{C}_D$ is almost optimal according to \cite{G2}.
 Furthermore, we can see that  $\mathcal{C}_D$ is an LCD distance optimal code according to the tables in (\cite{GK, HS}),

 }
\end{example}


\section{Concluding remarks}

In this paper we obtain an infinite family of binary LCD codes and two infinite families
of binary self-orthogonal codes by using simplicial complexes.
Weight distributions are explicitly determined for these codes.
We also find some (almost) optimal binary self-orthogonal and LCD codes.

It is worth noting that some of our self-orthogonal codes in Theorem 3.1  meet the Griesmer bound.
Table I presents some of optimal binary LCD codes obtained by using Theorems 3.6 and 3.7; their optimality is based on the tables in (\cite{GK, HS}). Their classification in (\cite{GK, HS}) treats binary LCD codes of only small lengths, so that optimality of our LCD codes in Theorems 3.6 and 3.7 is confirmed for only small lengths due to limited current database.
However, we believe that our binary LCD codes may include new LCD distance optimal codes of larger lengths provided that the database is supported for larger lengths.

\begin{table} [h]
\caption{some of LCD distance (or almost) optimal codes from Theorems 3.6 and 3.7 }
\begin{tabu} to 0.4\textwidth{|X[1,c]|X[1,c]|}
\hline
\rm{Parameters}&\rm{Optimality} \\
\hline
$[3,2,2]$ & LCD distance optimal \\
\hline
$[6,4,2]$ & LCD distance optimal \\
\hline
$[6,2,3]$ & LCD distance optimal \\
\hline
$[9,3,3]$ & LCD almost optimal \\
\hline
$[9,6,2]$ & LCD distance optimal \\
\hline
$[12,8,2]$ & LCD distance optimal \\
\hline
$[15,10,2]$ & LCD almost optimal \\
\hline
\end{tabu}
\end{table}

{\bf Acknowledgement.}
We express our gratitude to the reviewers for their very helpful comments, which
improved the exposition of this paper.





%


\ifCLASSOPTIONcaptionsoff
  \newpage
\fi


\begin{thebibliography}{1}

\bibitem{BCCGM} J. Bringer, C. Carlet, H. Chabanne, S. Guilley,  H. Maghrebi, Orthogonal direct sum masking, a smartcard friendly computation paradigm in a code, with builtin protection against side-channel and
fault attacks, in Proc. WISTP, 40-56, 2014.

\bibitem{CRSS1} A. K. Calderbank, E. M. Rains, P. W. Shor,  N. J. A. Sloane,
Quantum error correction and orthogonal geometry, Phys. Rev. Lett, 78(3-20): 405-408, 1997.

\bibitem{CRSS2}  A. R. Calderbank, E. M. Rains, P. W. Shor, and N. J. A. Sloane, Quantum error correction via codes over $GF(4)$, IEEE Trans. Inf.
Theory, 44(4): 1369-1387, 1998.


\bibitem{CG}  C. Carlet, S. Guilley, Complementary dual codes for counter-measures to
side-channel attacks, in Coding Theory and Applications (CIM Series
in Mathematical Sciences), vol. 3, E. R. Pinto, Ed. Berlin, Germany:
Springer-Verlag,  97-105, 2014.



  \bibitem{CMTQ1} C.  Carlet, S. Mesnager, C. Tang, Y. Qi, New characterization and parametrization of LCD codes, IEEE Trans. Inf. Theory, 65(1): 39-49, 2019.






 \bibitem{CMTQ2} C. Carlet, S. Mesnager, C. Tang, Y. Qi, Euclidean and Hermitian LCD MDS codes,  Des. Codes Cryptogr.,  86(11): 2605-2618, 2018.



 \bibitem{CMTQ3} C. Carlet, S. Mesnager, C. Tang, Y. Qi, On $\sigma$-LCD codes, 65(3):1694-1704, 2019.

\bibitem{CMTQP} C. Carlet, S. Mesnager, C. Tang, Y. Qi,  R. Pellikaan, Linear
codes over $\Bbb F_q$ are equivalent to LCD codes for $q > 3$,  IEEE  Trans. Inform. Theory, 64(4):  3010-3017, 2019.


\bibitem{CH} S. Chang,  J. Y. Hyun,  Linear codes from simplicial complexes,  Des. Codes Cryptogr.,  86:  2167-2181, 2018.






\bibitem{GK} L. Galvez, J-L. Kim, N. Lee, Y. G. Roe, B-S, Won, Some bounds on binary LCD codes, Cryptogr. Commun., 10(4): 719-728, 2018.

\bibitem{G2} M. Grassl,  Bounds on the minimum distance of linear codes.  http://www.codetables.de.




 \bibitem{G} J. H. Griesmer, A bound for error correcting codes, IBM J. Res. Dev., 4: 532-542, 1960.





\bibitem{HS} M. Harada, K. Saito, Binary linear complementary dual codes, Cryptogr. Commun., 11(4): 677-696, 2019.



\bibitem{HP}  W. C. Huffman, V. Pless, Fundamentals of Error-Correcting Codes, Cambridge University Press, Cambridge, 2003.

 \bibitem{KY}  X. Kong, S. Yang,  Complete weight enumerators of a class of linear codes with two or three weights, Discrete Math. 342: 3166-3176, 2019.

 \bibitem{L} C. Li, Hermitian LCD codes from cyclic codes, Des. Codes Cryptogr., 86(10): 2261-2278, 2018.

 \bibitem{LDL}  C. Li, C. Ding,  S. Li, LCD cyclic codes over finite fields, IEEE
Trans. Inf. Theory,  63(7):  4344-4356,  2017.
















































































































































\bibitem{M}  J. L. Massey, Linear codes with complementary duals, Discrete Math., volumes 106-107: 337-342,  1992.






















 \bibitem{Y}  S. Yang, Q. Yue, Y. Wu, X. Kong, Complete weight enumerators of a class of two-weight linear codes. Cryptogr.
Commun., 11:   609-620, 2019.


 \bibitem{ZTLD}  Z. Zhou, C. Tang, X.  Li, C. Ding,  Binary LCD codes and self-orthogonal codes from a generic construction, IEEE Trans. Inf. Theory, 65(1): 16-27, 2019.


























\end{thebibliography}
\end{document}